\def\dddot#1{{\hspace{3pt}\dot{\phantom #1}\hspace{-2.4pt}\ddot{\!#1}}}

\documentstyle[eqsecnum,aps]{revtex}
\def\Der{D}

\def\ins#1{}

\def\iems#1{}
\def\comment#1{}

\def\cm#1{}

\begin{document}
\setcounter{figure}{0}
\Roman{figure}

\title{
Covariant Effective Action for Quantum Particle
with Coordinate-Dependent Mass
}
\author{
 H. Kleinert%
 \thanks{Email: kleinert@physik.fu-berlin.de 
URL:
http://www.physik.fu-berlin.de/\~{}kleinert \hfill
}
 and A.~Chervyakov\thanks{chervyak@physik.fu-berlin.de\newline
 On leave from LCTA, JINR, Dubna, Russia} }
\address{Institut f\"ur Theoretische Physik,\\
Freie Universit\"at Berlin, Arnimallee 14,
14195 Berlin, Germany}
\maketitle
\begin{abstract}
Using a covariant background field method
we calculate the one-loop quantum effective action
for a particle with coordinate-dependent mass
moving slowly through a one-dimensional configuration space.
The procedure can easily be extended to any desired
loop order.
\end{abstract}

%
\section{Introduction}
Just as the classical behavior
of systems is completely determined
by the extrema of the
classical
action, the
quantum-mechanical properties
can be found from
the extrema
of
the so-called
effective action.
This can be calculated
 perturbatively
in powers of
the Planck constant
$\hbar $
from the one-particle-irreducible (1PI) Feynman diagrams
with no external
quantum lines. The leading order
is
the classical action itself,
whereas the one-loop term is a
trace of a logarithm,
 ${\rm Tr\log} M$, where $M$
is a
functional matrix associated with small quantum fluctuations.
The rules for calculating higher orders are well known.

Recently \cite{jl}, the one-loop
effective action was calculated for
a particle in one dimension.
Both the potential and the kinetic
terms are modified by terms of order
$\hbar$. The result  is
 not applicable
to a large variety
of interesting physical systems,
for instance compound nuclei, where
the collective Hamiltonian, commonly
derived from a microscopic description
via
 time-dependent
Hartree-Fock theory \cite{gr},
contains coordinate-dependent mass parameters.
This requires the extension of the effective
action formalism to reparametrization-invariant
quantum-mechanical systems in a curved spacetime
similar to
 one-dimensional nonlinear $ \sigma $-models.

A straightforward tool
for computing the effective action
of quantum-mechanical systems
is provided by the background field
method which in the context of
 quantum mechanics
is described in the textbook \cite{PI}.
In quantum field theory, this method
provides a way
to keep manifest the reparametrization
invariance of $ \sigma $-models, and is therefore
the preferred method for explicit
calculations of effective actions
 \cite{AGFM}.
In this paper we use this
 method to calculate the effective action
for a particle with coordinate-dependent mass.
It is the simplest
reparametrization-invariant quantum-mechanical
$ \sigma $-model with
a one-dimensional parameter
space.
We generalize this model by including into
the classical action
a
scalar potential.
The
model is UV-finite---all
 intermediate divergences will be seen to cancel
against those
coming from the invariant measure \cite{KlCh}.

\section{Background field method}
Consider a particle with coordinate-dependent
mass $m(x)$ moving as in the one-dimensional potential
$V(x)$. We shall study the euclidean version
of the system where the paths $x(t)$ are
continued to an  imaginary times
$\tau =-it$ and the Lagrangian
for
 $x(\tau )$  has the form
\begin{equation}
 {\cal L} (x, \dot x) =
\frac{1}{2}\,m(x)\dot x ^2 + V(x).
\label{1}\end{equation}
The ${\rm dot}$ stands for the
 derivative with respect to the
imaginary time.
The $x$-dependent mass may be written as
$mg(x)$ where $g(x)$ plays the role
of a one-dimensional dynamical metric.
It is the trivial $1\times 1$  {\em Hessian matrix\/} of the system.
In $D$-dimensional configuration space,  the kinetic term would read
$mg_{\mu \nu }(x)x^\mu x^ \nu /2$
as in
an O($D$)-symmetric nonlinear $ \sigma $-model in one dimension.

Under
an arbitrary single-valued coordinate transformation
 $x = x(y)$, the
potential $V(x)$ is assumed to
transform like a scalar whereas
 the metric $m(x)$ is a one-dimensional
tensor of rank two:
\begin{eqnarray}
\!\!\!\!\!\!\!\!\!\!\!\!\!\!\!\!\!\!\!\!\!\!
 V(x) = V(x(y)) \equiv \widetilde V(y),~~~~~
 m(x) = \widetilde m(y)\,{y'}^2 (x).
\label{2}\end{eqnarray}
This coordinate transformation leaves the Lagrangian (\ref{1})
and thus also the
classical action
\begin{eqnarray}
 {\cal A} [x] = \int_{-\infty }^\infty  d \tau\,  {\cal L} (x, \dot x)
\label{3}\end{eqnarray}
invariant.

The quantum theory has to possess the same invariance.
In Schr\"odinger theory, this is
automatic. In the path integral
formulation of the quantum system,
this property is nontrivial, and only recently
the authors have shown how to proceed
to guarantee it.\cite{KlCh}
Of course, this invariance must also be manifest
in the effective action
of the  quantum theory.
This will be achieved
by combining
the  background field method\cite{AGFM}
with the techniques of Refs.~\cite{KlCh}.
In the
background field method
we split all paths into $x(\tau ) = X(\tau ) + \delta x(\tau )$,
where $X(\tau )$ is the final extremal orbit
and $\delta x$
describes the
quantum fluctuations around
it.
At the one-loop level,
the
 covariant effective
action $\Gamma [X]$ becomes
a sum of
the
 classical Lagrangian  ${\cal L}(X,\dot X)$
and a correction term
$  \Delta {\cal L }$.
It is defines by the path integral
\begin{eqnarray}
e^ {- \Gamma [X]/\hbar } = \int {\cal D } \mu (\delta x)
e^{-\frac{1}{\hbar}\left\{  {\cal A}  [X + \delta x]
   - \int d \tau\, \delta x \,\frac{\delta {\Gamma} [X]}{\delta X}\right\}}\,,
\label{4}\end{eqnarray}
where the measure of functional
 integration ${\cal D}\mu (\delta x)$
is obtained from the initial invariant
measure ${\cal D }\mu( x) = Z^{-1}\,\prod_{\tau}
 d  x (\tau) \, \sqrt{m (x)}\,$ and reads
\begin{eqnarray}
 {\cal D }\mu(\delta x) = Z^{-1}\,\prod_{\tau}
 d \delta x (\tau) \, \sqrt{m (X)}\,
 e ^{\frac{1}{2}  \delta (0) \int d \tau \log
     \frac{m(X + \delta x)}{m(X)}}\,,
\label{5}\end{eqnarray}
with $Z$ being some normalization factor.
The generating functional (\ref{4}) possesses the same symmetry under
configuration space reparametrizations as the classical action (\ref{3}).

We now calculate $\Gamma [X]$
in Eq.~(\ref{4}) perturbatively
as a power series in $ \hbar $
\begin{eqnarray}
 \Gamma [X]  = {\cal A} [X] + \hbar  \Gamma _1 [X] +
 \hbar^2  \Gamma _2 [X] + \dots\,.
\label{6}\end{eqnarray}
The quantum corrections to the classical action (\ref{3})
are obtained by expanding ${\cal A} [X + \delta x]$
and
the measure (\ref{5})
in powers of $\delta x$:
\begin{equation}
{\cal A} [X+\delta x]=
{\cal A} [X]+\int\! d\tau \frac{ D{\cal A}  }{ \delta X^\mu}\delta x^\mu
+\frac{1}{2} \int\! d\tau\!\!\int d\tau'  \frac{ D^2{\cal A}  }{
 \delta X^\mu
 \delta X^\nu
}        \delta x^\mu    \delta x^\nu
+\frac{1}{6} \!
\int\! d\tau \!\!
\int \!d\tau'\!\!\!
\int d\tau '' \!
  \frac{ D^3{\cal A}  }{
 \delta X^\mu
 \delta X^\nu
 \delta X^ \lambda
}        \delta x^\mu
  \delta x^\nu
  \delta x^\lambda  +\dots\,.
\label{@hbarex}\end{equation}
The symbols $D/ \delta X^\mu$ denote covariant
functional derivatives.
To first order, this is the ordinary derivative
\begin{eqnarray}
\frac{\Der  {\cal A} [X]}{\delta X (\tau)} =
\frac{\delta {\cal A} [X]}{\delta X (\tau)} =
V'(X) - \frac{1}{2}\,m' (X)\,{\dot X}^{2}(\tau) - m (X)\,{\ddot X} (\tau).
\label{7}\end{eqnarray}
This vanishes for the correct physical orbit $X(t)$.
The second covariant derivative is
\begin{eqnarray}
\frac{\Der  ^2 {\cal A} [X]}{\delta X (\tau) \delta X (\tau')} =
\frac{\delta ^2 {\cal A} [X]}{\delta X (\tau) \delta X (\tau')} -
  \gamma (X)\, \frac{\delta {\cal A} [X]}{\delta X (\tau)}\,\equiv
M(\tau ,\tau '),
\label{8}\end{eqnarray}
where $\gamma (X) = m' (X)/2\,m (X)$ is the Christoffel symbol
for the
metric $m (X)$, and
 and the  functional second derivative reads explicitly
\begin{eqnarray}
\frac{\delta ^2  {\cal A} [X]}{\delta X (\tau) \delta X (\tau')} =
- \left[m(X)\frac{d^2}{d\tau ^2} + m'(X)\dot X \frac{d}{d\tau}
   + m'(X)\ddot X + \frac{1}{2}m''(X) {\dot X}^2 - V'' (X)\right]
   \delta (\tau - \tau')\,.
\label{9}\end{eqnarray}
The validity of the
expansion (\ref{@hbarex}) follows from the fact that it is
equivalent by a coordinates transformation
to an ordinary functional expansion in Riemannian coordinates
where the Christoffel symbol vanishes
for the particular background coordinates.

The inverse of the
functional matrix $M(\tau ,\tau ')$ in
(\ref{8})
supplies us with
 the propagator  $G(\tau ,\tau ')\equiv M^{-1}(\tau ,\tau ')$.
The higher derivatives define the interactions.
The expansion terms
 $\Gamma_n [X]$
in (\ref{6})
are found from all one-particle-irreducible
Feynman diagrams formed with the
propagator $G(\tau ,\tau ')$
and the interaction vertices.
The one-loop correction to the effective action
is given
by the simple harmonic
path integral
\begin{eqnarray}
e^ {- \Gamma_1 [X] } = \int {\cal D }\delta x \, \sqrt{m (X)}\,
e^{- {\cal A}^{(2)} [X, \delta x]  }\,,
\label{10}\end{eqnarray}
with the quadratic part of the expansion
(\ref{@hbarex}):
\begin{eqnarray}
{\cal A}^{(2)} [X, \delta x] =
\frac{1}{2}\,\int _{-\infty }^\infty d\tau\,d\tau'\,\delta x (\tau)\,
\frac{\Der  ^2 {\cal A} [X]}{\delta X (\tau) \delta X (\tau')}\,
\delta x (\tau')\, .
\label{11}\end{eqnarray}

The presence of $m (X)$ in the free part
of the covariant kinetic term (\ref{11})
and in the measure in Eq.~(\ref{10})
suggests exchanging
the fluctuation
$\delta x$ the new coordinates
$\tilde x = h(X)\delta x$, where
$h(X)\equiv  \sqrt{m(X)}$ is the
einbein $h(X) = \sqrt{m(X)}$ associated with
the metric $m(X)$. Its covariant derivative is
$D_X h(X) = \partial_X h(X) - \gamma (X)\,h(X)\equiv 0$.
Then
(\ref{11}) becomes
\begin{eqnarray}
{\cal A}^{(2)} [X, \delta x] =
\frac{1}{2}\,\int _{-\infty }^\infty d\tau\,d\tau'\,
\tilde x (\tau)\,
\frac{\Der  ^2 {\cal A} [X]}{\delta {\tilde X} (\tau)
\delta {\tilde X} (\tau')}\,
\tilde x (\tau')\,,
\label{12}\end{eqnarray}
with
\begin{eqnarray}
\frac{\Der  ^2 {\cal A} [X]}{\delta {\tilde X} (\tau)
\delta {\tilde X} (\tau')} = e^{-1} (X)\,
\frac{\Der  ^2 {\cal A} [X]}{\delta X (\tau) \delta X (\tau')}\,
 e^{-1} (X) =
\left[ - \frac{d^2}{d\tau ^2} + \Omega ^2 (X)\right]
 \delta (\tau - \tau')\,,
\label{13}\end{eqnarray}
where
\begin{eqnarray}
\Omega ^2 (X)  =
 e^{-1}(X)\,\Der^2  V(X)\,e^{-1}(X) =
 e^{-1}(X)\,\Der  V'(X)\,e^{-1}(X) =
\frac{1}{m(X)}\,
  \left[V''(X) - \gamma(X)\,V'(X)\right]\equiv \tilde M(\tau ,\tau ').
\label{14}\end{eqnarray}
Thus the  fluctuations $\tilde x$ behave like
those of a harmonic oscillator with the
time-dependent frequency
 $\Omega^2 (X)$, which is obviously
a scalar under reparametrizations of configuration  space.
By construction, the fluctuation vanishes at the infinity
$\tilde x (\tau)\mathop{}_{|\tau|\rightarrow \infty} \longrightarrow 0$.
In the coordinates $\tilde x(\tau )$, the functional measure
of integration
 in Eq.~(\ref{10})
takes the euclidean
 form
\begin{eqnarray}
 \prod_{\tau} d \delta x (\tau)\sqrt{m(X)}  = \prod_{\tau} d \tilde x (\tau)\,.
\label{15}\end{eqnarray}
This allows us to
 integrate  the Gaussian path integral (\ref{10}) trivially
to obtain the one-loop quantum correction to the effective action
\begin{eqnarray}
 \Gamma_1 [X] = \frac{1}{2}{\rm Tr}\log\left[- d^2_{\tau} +
     \Omega ^2(\tau )\right]\,.
\label{16}\end{eqnarray}
This expression is
a nonlocal
functional of the background
field $X (\tau)$. It cannot be evaluated explicitly.
For sufficiently slow
movements of $X(\tau )$, however,
we can resort to a gradient expansion
which yields  asymptotically
a local expression for the effective action.

\section{Derivative expansion}

The derivative expansion of the one-loop
effective action (\ref{16}) has the general form
\begin{eqnarray}
\Gamma_1 [X] =  \int _{-\infty }^\infty d\tau\, \left[V_1 (X) +
\frac{1}{2}\,Z_1 (X)\,{\dot X}^2 + \cdots ~\right]\,,
\label{23}\end{eqnarray}
where $V_1 (X)$ is the one-loop effective
potential which can be found explictly
by calculating
the trace log in (\ref{16})
for a time-independent
$ \Omega ^2$:
\begin{eqnarray}
 V_1 (X) =
 \frac{1}{2}{\rm Tr}\log\left[- d^2_{\tau} +
     \Omega ^2(\tau )\right]\,=
 \frac{1}{2}\,\int
 \frac{d k}{2\pi}\,\log (k^2 +  \Omega ^2 ) = \frac{1}{2}\,\Omega (X)\,.
\label{28}\end{eqnarray}
By construction,  $V_1 (X)$ and  $Z_1 (X)$
transform like a scalar and a  tensor of rank 2
under reparametrizations of
configuration space.
In order to find $Z_1 (X)$
we
 split the background field as
$X(\tau) = X_0 + \delta X(\tau)$,
where $X_0$ is a {\em constant\/} field
and $\delta X(\tau)$ is slowly varying,
and expand
(\ref{16})
in powers of $\delta X$
up to ${\delta X}^2$.
The general
expression (\ref{28})
shows that the second order expansion
must have the form
\begin{eqnarray}
\Gamma_1 [X_0 + \delta X] - \Gamma_1 [X_0]
=  \int _{-\infty }^\infty d\tau\, \left[\Der  V_1 (X_0)\,\delta X
 + \frac{1}{2}\,\Der^2  V_1 (X_0)\,{\delta X}^2 +
 \frac{1}{2}\,Z_1 (X_0)\,\dot{\delta X} ^2 \right]\,,
\label{24}\end{eqnarray}
where
\begin{eqnarray}
\Gamma_1 [X_0] =
\frac{1}{2}\,{\rm Tr}\log\left[ -d^2_\tau + \Omega ^2 (X_0)\right]\,.
\label{24bis}\end{eqnarray}
Such an expression is indeed obtained
from the expansion of (\ref{16}) which yields
\begin{eqnarray}
\Gamma_1 [X_0 + \delta X] - \Gamma_1 [X_0]
=  \frac{1}{2}\,{\rm Tr}\log\left\{ 1 + \frac{1}{\hat k^2 + \Omega ^2 }\,
  \left[\Der  \Omega^2 (X_0)\,\delta X +
\frac{1}{2}\,\Der ^2  \Omega^2 (X_0)\,{(\delta X)} ^2\right] \right\}\,,
\label{25}\end{eqnarray}
where
$ \Omega ^2\equiv  \Omega ^2(X_0)$ and
 we have introduced  the wave number operator  $\hat k \equiv  -id_\tau$.
Expanding the logarithm in Eq.~(\ref{25})
up to ${\delta X}^2$, we obtain
\begin{eqnarray}
\Gamma_1 [X_0 + \delta X] - \Gamma_1 [X_0]\,& = &
    \frac{1}{2}\,{\rm Tr}\left\{ \frac{1}{\hat k^2 + \Omega ^2 }\,
 \left[\Der  \Omega^2 (X_0)\,\delta X +
 \frac{1}{2}\,\Der ^2  \Omega^2 (X_0)\,{(\delta X)} ^2\right] \right\}
\nonumber\\
&-&\, \frac{1}{4}\,(\Der  \Omega^2)^2 (X_0) \,
{\rm Tr}\left[ \frac{1}{\hat k^2 + \Omega ^2 }\,\delta X \,
   \frac{1}{\hat k^2 + \Omega ^2 }\,\delta X \right] \,.
\label{26}\end{eqnarray}
The term linear in $\delta X$ can immediately be calculated in the spectral
representation
of the propagator:
\begin{eqnarray}
    \frac{1}{2}\,\Der  \Omega^2 (X_0)\,\int_{-\infty }^\infty d \tau\, \delta X
(\tau)\,\int
 \frac{d  k}{2\pi}\,\frac{1}{ k^2 + \Omega ^2 }\,=
    \frac{1}{4\Omega (X_0)}\,\Der  \Omega^2 (X_0)\int_{-\infty }^\infty d
\tau\, \delta X (\tau)\,
{}.
\label{27}\end{eqnarray}
We now recall the zeroth order term in Eq.~(\ref{28})
and rewrite this as
\begin{eqnarray}
 \int_{-\infty }^\infty d \tau \,\Der  V_1 (X_0)\delta X (\tau)\,,
 \label{29}\end{eqnarray}
thus reproducing the
 the linear $\delta X (\tau)$-term in the expansion (\ref{24}).
In the calculation it is important that
both $V_1 (X)$ and $\Omega^2 (X)$
are the scalars under reparametrizations, such that
the first covariant derivatives
are equal to the ordinary derivatives:
$\Der  V_1 (X) =d V_1 (X)/dX$,\,
 $\Der  \Omega^2 (X) =d \Omega^2 (X)/dX$.
The scalar
%
%
nature of $ \Omega ^2(X)$  is
obvious from the following
 covariant
representation
\begin{eqnarray}
\Omega ^2 (X)  =
\frac{1}{\sqrt{m(X)}}\,\frac{d}{d X}\,
  \left[\sqrt{m(X)}\left(\frac{V'(X)}{m(X)}\right)\right]\,,
\label{30bis}\end{eqnarray}
which is the one-dimensional version
of the Laplace-Beltrami expression
$g^{-1/2}\partial _\mu g^{1/2}g^{\mu \nu }\partial _ \nu V(X).$

The quadratic terms in $\delta X$
in Eq.~(\ref{26}) read
\begin{eqnarray}
\frac{1}{4}\,\Der^2  \Omega^2 (X_0)\,{\rm Tr}\left[\frac{1}{
\hat k^2 + \Omega ^2 }\,
  {(\delta X)} ^2\right]
 - \frac{1}{4}\,(\Der  \Omega^2)^2 (X_0)\,
{\rm Tr}\left[ \frac{1}{\hat k^2 + \Omega ^2 }\,\delta X \,
   \frac{1}{\hat k^2 + \Omega ^2 }\,\delta X \right] \,.
\label{31}\end{eqnarray}
The first functional trace is immediately calculated
with the spectral representation of the propagator
and yields
\begin{eqnarray}
\frac{1}{4}\,\Der^2  \Omega^2 (X_0)\,
\int\,\frac{d k}{2\pi}\,\frac{1}{k^2 + \Omega ^2}\,
= \frac{1}{2}\,\frac{\Der ^2  \Omega^2 (X_0)}{4\,\Omega (X_0)}
\int_{-\infty }^\infty  d \tau\,{(\delta X)}^2(\tau )\,.
\label{35bis1}\end{eqnarray}
The evaluation of the second term
in (\ref{31})
needs more work
due to the $\tau $-dependence of $ \delta X(\tau )$
between the differential operators.
In order to use the spectral representation,
we must
move
all operators $\hat k$ to the left
of all $ \delta X(\tau )$.
Making use of the
identity\cite{Fr}
\begin{eqnarray}
 \delta X \,\frac{1}{\hat k^2 + \Omega ^2} =
 \frac{1}{\hat k^2 + \Omega ^2}\,\delta X +
 \frac{1}{(\hat k^2 + \Omega ^2)^2}\,
 \left[\hat k^2,\,\delta X \right] +
 \frac{1}{(\hat k^2 + \Omega ^2)^3}\,
 \left[\hat k^2,\,\left[\hat k^2,\,\delta X \right]\right] + \cdots
\label{32}\end{eqnarray}
and the obvious commutation relations
\begin{eqnarray}
 \left[\hat k^2,\,\delta X \right] =
 \delta  \ddot{ X}  - 2i\hat k \delta  \dot{ X}
\label{33}\end{eqnarray}
and
\def\dddot#1{\!\!\hspace{3pt}\dot{\phantom
#1\!\!\!}\hspace{-1.6pt}\ddot{\!\!#1}}
\def\ddddot#1{\,{\ddot{\!\ddot#1}}}
\begin{eqnarray}
 \left[\hat k^2,\,\left[\hat k^2,\,\delta X \right]\right] =
 \delta \ddddot X
-4i\hat k \delta \dddot X
-4\hat k^2 \delta \ddot X ,
\label{34}\end{eqnarray}
we reduce the second term in
(\ref{31}) to the following form
\begin{eqnarray}
- \frac{1}{4}\,(\Der  \Omega^2)^2 (X_0)\,
{\rm Tr}\left[ \frac{1}{(\hat k^2 + \Omega ^2)^2}\,{(\delta X)}^2 \right]
- \frac{1}{4}\,(\Der  \Omega^2)^2 (X_0)\,
 {\rm Tr}\left\{\left[\frac{1}{(\hat k^2 + \Omega ^2)^3}
- \frac{4\hat k^2}{(\hat k^2 + \Omega ^2)^4}\right]\,\delta X \delta \ddot{
X}\right\}\,,
\label{35}\end{eqnarray}
where the terms containing higher time derivatives
of $\delta X (\tau)$ have to be omitted.
Here the first term gives  immediately
\begin{eqnarray}
-\frac{1}{4}\,(\Der  \Omega^2)^2 (X_0)\,\int\,\frac{d k}{2\pi}\,
\frac{1}{(k^2 + \Omega ^2)^2}\,\int d \tau\,{(\delta X)}^2 (\tau )
= -\frac{1}{2}\,
\frac{(\Der  \Omega^2)^2 (X_0)}{8\,{\Omega}^3 (X_0)}
\int_{-\infty}^\infty  d \tau\,{(\delta X)}^2(\tau )\,.
\label{35bis}\end{eqnarray}
Combining this with
(\ref{35bis}), we find
the quadratic term in $ \delta X(\tau )$:
\begin{equation}
  \frac{1}{2}\,\Der^2  V_1 (X_0)\,
 \int _{-\infty }^\infty d\tau\,
{\delta X}^2(\tau ),
\label{@}\end{equation}
as follows directly
from the definition
$\Der^2   V_1 (X_0) = V''_1 (X_0)
 - \gamma (X_0)\,V'_1 (X_0)=
$ of Eq.~(\ref{24}) and $ V_1 (X_0)= \Omega (X_0)/2$.

The last term in (\ref{35}) is
\begin{eqnarray}
- \frac{1}{4}\,(\Der  \Omega^2)^2 (X_0)\,
\,\int\,\frac{d k}{2\pi}\,\frac{\Omega^2 - 3k^2}{(k^2 + \Omega^2)^4 }
 \int _{-\infty}^\infty d \tau \, \delta X (\tau) \, \delta \ddot{ X}
 (\tau )
=
-
\frac{(\Der  \Omega^2)^2 (X_0)}{64\,\Omega ^{5} (X_0)}\,\int d \tau
   \,
 \delta X (\tau) \, \delta \ddot{ X}
 (\tau )
{}.
\label{36}\end{eqnarray}
After a partial integration in $\tau $, this can be compared to
 last term in the gradient expansion
(\ref{24}).
with the result
\begin{eqnarray}
Z_1 (X_0) =
 \frac{(\Der  \Omega^2)^2 (X_0)}{32\,\Omega ^5 (X_0)}\,.
\label{39}\end{eqnarray}
We may now replace $X_0$ by $X (\tau)$
to obtain, finally, the one-loop correction
to the kinetic term
\begin{eqnarray}
Z_1 (X) =
 \frac{(\Der \Omega)^2 (X)}{8\,\Omega ^3 (X)}\,.
\label{40}\end{eqnarray}
Substituting this into
with (\ref{28}) into
(\ref{23})  and this into
the expansion (\ref{6})
we obtain
 effective action
to order $\hbar$:
\begin{eqnarray}
\Gamma^{\rm eff} [X] = \int _{-\infty }^\infty d\tau\,  \left[
\frac{1}{2}\,m^{\rm eff} (X)\,{\dot X}^2 +
 V^{\rm eff} (X) \right]\,,
\label{41}\end{eqnarray}
where the bare metric $m^{\rm eff}$ and the potential
$V^{\rm eff}$ differ from the initial classical expressions
by the above one-loop fluctuation corrections:
\begin{eqnarray}
 m^{\rm eff} (X) &=& m(X) + \hbar \,
 \frac{(\Der  \Omega)^2 (X)}{8\,\Omega ^3 (X)}\,, \nonumber\\
 V^{\rm eff} (X) &=& V(X) + \hbar\,\frac{1}{2}\,\Omega (X)\,.
\label{42}\end{eqnarray}
The effective action
has the same
coordinate independence
as the initial action. For an $x$-independent mass, the result reduces to that
of
 Ref.~\cite{jl}.

A final word is necessary on the range of validity of the expansion.
The derivation shows that
the characteristic time scale is $1/ \Omega $.
Within this time, the particle has to move
only little, i,e., $\dot X(\tau )/X(\tau )$ has top be much smaller than $
\Omega $.

\section{Conclusion}
We have derived
the one-loop approximation to the
covariant
 effective
action for a point particle with
coordinate-dependent mass moving slowly
through a  one-dimensional scalar potential.
The extremization of this action yields an equation of motion whose
solutions contain all
quantum effects linear in $\hbar $.



\begin{thebibliography}{11}
%
\bibitem{jl}
F.~Cametti, G.~Jona-Lasinio, C.~Presilla and F.~Toninelli,
{\em Comparison between quantum and classical dynamics in the
effective action formalism,\/}
Proc. of the Int. School of Physics ''Enrico Fermi'',
CXLIII Ed. by G.~Casati, I.~Guarneri, U.~Smilansky,
Amsterdam, IOS Press, 2000, pp. 431-448
{\tt quant-ph/9910065};
B.R.~Friedan and A.~Plastino,
{\em Classical trajectories compatible with quantum mechanics,\/}\\
quant-ph/0006012 (2000).
%
\bibitem{gr}
K.~Goeke and P.-G.~Reinhart,
Ann. Phys. {\bf 112} (1978) 328.
%
\bibitem{PI}
H.~Kleinert,
{\em Path Integrals in Quantum Mechanics, Statistics and Polymer Physics,\/} \\
World Scientific Publishing Co., Singapore 1995,
Second extended edition. For the original literature see
G.~'t Hooft, Nucl.~Phys.~{\bf B62} (1973) 444.
%
\bibitem{AGFM}
L.~Alvarez-Gaum\'e, D.Z.~Freedman and S.~Mukhi, Ann.\
of Phys.~{\bf 134} (1981) 85;\\
see also:\\
%
J.~Honerkamp, Nucl.~Phys.~ {\bf B36} (1972) 130;\\
%
G.~Ecker and J.~Honerkamp, Nucl.~Phys.\
{\bf B35} (1971) 481;\\
%
L.~Tataru, Phys.~Rev.~{\bf D12} (1975) 3351;\\
%
G.~A.~Vilkoviski, Nucl.~Phys.~{\bf B234} (1984) 125;\\
%
E.S.~Fradkin and A.A.~ Tseytlin, Nucl.~Phys.~{\bf B234} (1984)
 509;\\
%
E.~Braaten, T.~L.~Curtright and C.~K.~Zachos,
Nucl.~Phys.~{\bf B260} (1985) 630;\\
%
P~.S.~Howe, G.~Papadopoulos and K.~S.~Stelle,
Nucl.~Phys.~{\bf B296} (1988)
26;\\
V.~V.~Belokurov and D.~I.~Kazakov,
Particles \& Nuclei ~{\bf 23} (1992) 1322.
%
\bibitem{KlCh}
H.~Kleinert and A.~Chervyakov, Phys.~Lett.~{\bf B464} (1999) 257;\\
hep-th/9906156;\\
H.~Kleinert and  A.~Chervaykov, Phys.~Lett.~
{\bf B477} (2000) 373;\\ quant-ph/9912056.
%
\bibitem{Fr}
C.~M.~Fraser, Z.~Phys.~{\bf C28} (1985) 101;
see also:\\
%
J.~Iliopoulos, C.~Itzykson, A.~Martin,
Rev.~Mod.~Phys.~{\bf 47} (1975) 165;\\
%
K.~Kikkawa, Prog.~Theor.~Phys.~{\bf 56} (1976) 947;\\
%
H.~Kleinert, Fortschr. Phys.~{\bf 26} (1978) 565;\\
%
R.~MacKenzie, F.~Wilczek and  A.~Zee,
Phys.~Rev.~Lett.~{\bf 53} (1984) 2203;\\
%
I.~J.~R.~Aitchison and C.~M.~Fraser, Phys.~Lett.~{\bf B146} (1984) 63.
\end{thebibliography}
\end{document}